\documentclass[aps,prb,twocolumn,superscriptaddress,showpacs,amssymb,amsmath,floatfix]{revtex4-1}

\usepackage{graphicx}
\usepackage{hyperref}

\newcommand{\new}[1]{#1}

\begin{document}

\title{Noise calculations within the second-order von Neumann approach}

\author{Philipp Zedler}
\affiliation{Institut f{\"u}r Theoretische Physik, Technische Universit{\"a}t Berlin, D-10623 Berlin, Germany}

\author{Clive Emary}
\affiliation{Institut f{\"u}r Theoretische Physik, Technische Universit{\"a}t Berlin, D-10623 Berlin, Germany}

\author{Tobias Brandes}
\affiliation{Institut f{\"u}r Theoretische Physik, Technische Universit{\"a}t Berlin, D-10623 Berlin, Germany}

\author{Tom\'{a}\v{s} Novotn\'y}
\email{tno@karlov.mff.cuni.cz}
\affiliation{Department of Condensed Matter Physics, Faculty of
Mathematics and Physics, Charles University in Prague, Ke Karlovu 5,
CZ-12116 Praha 2, Czech Republic}

\begin{abstract}
We extend the second-order von Neumann approach within the generalized master equation formalism for quantum electronic transport to include the counting field. 
The resulting non-Markovian evolution equation for the reduced density matrix of the system resolved with respect to the number of transported charges enables the evaluation of the noise 
and higher-order cumulants of the full counting statistics. We apply this formalism to an analytically solvable model of a single-level quantum dot coupled to highly biased leads with 
Lorentzian energy-dependent tunnel coupling and demonstrate that, although reproducing exactly the mean current, the resonant tunneling approximation is not exact for the noise and higher order cumulants. Even if it may  
fail in the regime of strongly non-Markovian dynamics, this approach generically improves results of lower-order and/or Markovian approaches. 
\end{abstract}

\date{\today}
 \maketitle

{\em Introduction.} --- Although electronic transport through interacting nanostructures  has been a subject of intense study in the past decade, its theoretical description in general situations remains a challenging task. Standard complementary theoretical approaches are the Landauer-B\"{u}ttiker scattering formalism in the limit of negligible Coulomb correlations and the master equation for weak coupling to the leads, respectively.\cite{DattaBook1995,Bruus04} If both the single-particle coherence effects due to coupling to leads and many-body Coulomb correlations are important, highly complex phenomena can occur, such as, e.g., the Kondo effect at low temperatures.\cite{Bruus04} Even well above the Kondo temperature the interplay between coherence and correlations may lead to nontrivial consequences,\cite{FAB} which are hard to describe theoretically. 

The second-order von Neumann approach (2vN) based on decoupling of equations of motion for the density matrix,\cite{Pedersen05,*PedersenPhD08,Pedersen10} like its 
relative
the resonant tunneling approximation (RTA) based on a suitable infinite resummation of the perturbation series for the generalized master equation (GME),\cite{Schoeller94,Konig:PRL96,*KoenigPhD98} is capable of capturing such effects, as has been confirmed also by the successful description of pertinent experiments.\cite{Kubala:PRB06, Nilsson10} In Ref.~\onlinecite{Kubala:PRB06} the line-widths of the nonlinear conductance for a quantum dot asymmetrically coupled to leads in the Coulomb blockade regime were measured for both voltage-bias polarities and fully explained by the RTA, while in Refs.~\onlinecite{Nilsson10,Karlstrom:PRB11} a canyon of suppressed linear conductance through an InSb nanowire quantum dot was correctly \new{reproduced} by the 2vN as a function of the gate voltage and magnetic field. Furthermore, in a series of works,\cite{Pedersen07,FAB,Pedersen10} the 2vN has been applied to the single resonant level as well as double quantum dot models, finding that corrections beyond the lowest-order sequential tunneling (cotunneling) were described very well and conjecturing that the method yields exact solutions for non-interacting models.\cite{Pedersen10} This is a nontrivial feature as the method stems from the atomic limit (described by GMEs) complementary to the non-interacting case. 

Altogether, 2vN/RTA is a powerful and successful approximation and it is, therefore, natural to ask whether it can be extended to the calculation of other transport quantities such as noise and full counting statistics. This is exactly the subject of this \new{Brief Report} --- we extend the second-order von Neumann approach to the evaluation of the full counting statistics, and noise in particular, by incorporating the counting field in the theory and then test the performance of this generalization for the simplest possible model of a single-level quantum dot. We find that contrary to the stationary mean current, which is captured exactly by the 2vN, the noise and higher-order cumulants constituting the full counting statistics are not exact.  Nevertheless, it does yield significant improvement over simpler, more standard approximations.                    

{\em Generalized second-order von Neumann approach.} --- In our generalization of the second-order von Neumann scheme we use the equations of motion approach 
and  closely follow the corresponding derivation without the counting field in Ref.~\onlinecite{Pedersen05}. We only summarize the main steps and associated approximations and 
show the final results. A fully detailed derivation can be followed in Ref.~\onlinecite{ZedlerPhD11}. We consider a generic transport Hamiltonian\cite{Pedersen05} consisting of three parts,
$H\equiv H_{\mathrm{system}}+H_{\mathrm{leads}}+H_{T}$, with the central system (``dot'') Hamiltonian $H_{\mathrm{system}}\equiv\sum_a E_a |a\rangle\langle a|$ expressed in terms of 
the system many-body eigenstates $|a\rangle$ and eigenenergies $E_a$, standard noninteracting free-electron Hamiltonian for the two mutually biased ($\mu_{L}=eV/2,\ \mu_{R}=-eV/2$) leads  $H_{\mathrm{leads}}\equiv
\sum_{k,\alpha=L, R} E_{k\alpha} c^\dagger_{k\alpha} c_{k\alpha}$, and the tunneling Hamiltonian $H_{T} \equiv\sum_{k\alpha,ab}\big[ T^{*}_{ba}(k\alpha) c^
\dagger_{k\alpha}|a\rangle\langle b| +  T_{ba}(k\alpha)|b\rangle\langle a| c_{k\alpha}\big]$.

Using the standard prescription,\cite{Braggio06} we can write down the extended Liouville-von-Neumann equation for the generalized (non-Hermitian) density matrix $\rho(t,\chi)=\sum_{\{N_{\alpha}\}}\rho(t,\{N_{\alpha}\})e^{i\sum_{\alpha}N_{\alpha}\chi_{\alpha}},\ \rho^{\dag}(t,\chi)=\rho(t,-\chi),\ \chi\equiv(\chi_{L},\chi_{R})$ involving the counting fields $\chi_{\alpha}$ dual to the number $N_{\alpha}$ of charges passed through junction $\alpha$ (left or right) during time $t$. This equation reads (we set $\hbar\equiv 1$ and $e\equiv 1$ throughout the rest of the paper) 
\begin{equation}
i \frac{d}{dt}\rho(t,\chi)
  = H^+(\chi) \rho(t,\chi) - \rho(t,\chi)H^-(\chi),
  \label{LiouvilleVonNeumann}
\end{equation}
where the $\chi$-dependent Hamiltonian is generated by  modifying its tunnel part as follows: $H^\pm(\chi) \equiv H_{\mathrm{system}} + H_{\mathrm{leads}} + \sum_{k\alpha,ab}\big[
		\bar T_{ba}^\pm(k\alpha,\chi) c^\dagger_{k\alpha}|a\rangle\langle b| +  T_{ba}^\pm(k\alpha,\chi)|b\rangle\langle a| c_{k\alpha} \big]$, 
with\cite{Braggio06} $\bar T_{ba}^\pm(k\alpha,\chi) \equiv e^{\pm i\chi_\alpha/2}T^{*}_{ba}(k\alpha)$ and $\ T_{ba}^\pm(k\alpha,\chi) \equiv e^{\mp i\chi_\alpha/2}T_{ba}(k\alpha)$.

For states of the whole system plus leads, we choose a basis of tensor products $|ag\rangle\equiv|a\rangle \otimes |g\rangle$, with $|g\rangle$ a many particle
state of the leads.  We first evaluate the time-evolution of system matrix elements $w_{ab}(t, \chi)\equiv \sum_g\langle ag|\rho(t,\chi)|bg\rangle$ and generalized system-lead coherences $\phi_{ab}(t,\chi;k\alpha)\equiv \sum_g\langle ag|c^\dagger_{k\alpha}\rho(t,\chi)|bg\rangle$ to which $w_{ab}(t, \chi)$ directly couple. We then continue for the matrix elements of $c_{k\alpha}^\dagger c_{k'\alpha'}\rho(t,\chi)$ and $c_{k\alpha}^\dagger c^\dagger_{k'\alpha'}\rho(t,\chi)$. At this level, we apply factorization and truncation conditions analogous to Ref.~\onlinecite{Pedersen05} and, corresponding to the resonant tunneling approximation of Ref.~\onlinecite{KoenigPhD98}, keep ``non-diagonal matrix elements of the total density matrix up to the difference of one electron-hole pair in the leads'' --- i.e., for example, $\sum_g\langle ag|c^\dagger_{k\alpha}c_{k\alpha}c^\dagger_{k'\alpha'}\rho(t,\chi)|bg\rangle\approx f_{k\alpha}\sum_g\langle ag|c^\dagger_{k'\alpha'}\rho(t,\chi)|bg\rangle$ for $k\alpha\not=k'\alpha'$, $\langle ag|c_{k_1\alpha_1}c^\dagger_{k_2\alpha_2}c_{k_3\alpha_3}\rho(t,\chi)|bg\rangle\approx 0$, for $k_1\alpha_1$, $k_2\alpha_2$, and $k_3\alpha_3$ all different. We obtain a closed (though infinite) set of linear equations of motion for the reduced density matrix $w_{ab}(t, \chi)$ and coherences $\phi_{ab}(t,\chi;k\alpha)$, which are more conveniently expressed in the Laplace picture ($w_{ab}(z, \chi)=\int_{0}^{\infty}dt e^{-zt} w_{ab}(t, \chi)$ and analogously for $\phi_{ab}(z,\chi;k\alpha)$).  Assuming $\phi_{ab}(t=0,\chi;k\alpha)=0$ and using the abbreviations $T^\pm_{ba}\equiv T^\pm_{ba}(k\alpha,\chi)$, $\phi_{ab}\equiv\phi_{ab}(z,\chi;k\alpha)$, $\bar\phi_{ab}\equiv\phi_{ab}^*(\bar z,-\chi;k\alpha)$, ${T'}_{ab}^\pm\equiv T^\pm_{ab}(k'\alpha',\chi)$ and analogously for $\phi_{ab}',\ \bar\phi_{ab}',\ \bar T_{ab}$, and $\bar T'_{ab}$, our equations of motion read
\begin{subequations}
\begin{equation}\label{generalPWw}
\begin{split}
(&i z-E_a+E_b) w_{ab}(z,\chi)-i w_{ab}(t=0,\chi)\\
& = 
  \sum_{a',k\alpha}\left[
    T^+_{aa'}\bar\phi_{ba'}
  + \bar T^+_{a'a}\phi_{a'b} \right]
  -
  \sum_{b',k\alpha}\left[
    T^-_{b'b}\bar\phi_{b'a}
  + \bar T^-_{bb'}\phi_{ab'}\right],
\end{split}
\end{equation}

\begin{widetext}
\begin{equation}\label{generalPWphi}
\begin{split}
(&i z-E_a+E_b+E_k)\phi_{ab}  =  \sum_{a'}T^+_{aa'}f_{k\alpha} w_{a'b}(z,\chi)-\sum_{b'}T^-_{b'b}(1-f_{k\alpha})w_{ab'}(z,\chi) \\  
&+ \sum_{a',k'\alpha'\not=k\alpha} {T'}^+_{aa'} \frac{
\sum_{a''}\left[\bar T^+_{a''a'}(1-f_{k'\alpha'})
           \phi_{a''b} - T^+_{a'a''}f_{k\alpha}\bar\phi_{ba''}'\right] 
+\sum_{b'}\left[\bar {T'}^-_{bb'}f_{k'\alpha'}\phi_{a'b'}
    - T^-_{b'b}(1-f_{k\alpha})\bar\phi'_{b'a'}\right]
}{i z-E_{a'}+E_b+E_{k\alpha}-E_{k'\alpha'}}
\\
&+ \sum_{a',k'\alpha'\not=k\alpha}\bar T^+_{a'a}
\frac{
\sum_{a''}\left[{T'}^+_{a'a''}f_{k'\alpha'}
          \phi_{a''b} - T^+_{a'a''}f_{k\alpha}\phi_{a''b}\right] 
+\sum_{b'}\left[{T'}^-_{b'b}(1-f_{k'\alpha'})\phi_{a'b'}
    - T^-_{b'b}(1-f_{k\alpha})\phi'_{a'b'}\right]
}{i z-E_{a'}+E_b+E_{k\alpha}+E_{k'\alpha'}}\\
&+ \sum_{b',k'\alpha'\not=k\alpha}
T^-_{b'b}
\frac{
\sum_{a'}\left[\bar {T'}^+_{a'a}(1-f_{k'\alpha'})
          \phi_{a'b'} - T^+_{aa'}f_{k\alpha}\bar\phi'_{b'a'}\right]
+\sum_{b''}\left[\bar {T'}^-_{b'b''}f_{k'\alpha'}\phi_{ab''}    
    -T^-_{b''b'}(1-f_{k\alpha})\bar\phi'_{b''a}\right]
}{i z-E_a+E_{b'}+E_{k\alpha}-E_{k'\alpha'}} \\
&+  \sum_{b',k'\alpha'\not=k\alpha}
\bar T^-_{bb'}
\frac{
\sum_{a'}\left[{T'}^+_{aa'}f_{k'\alpha'}
         \phi_{a'b'} - T^+_{aa'}f_{k\alpha}\phi'_{a'b'}\right]
+\sum_{b''}\left[{T'}^-_{b''b'}(1-f_{k'\alpha'})\phi_{ab''}
    - T^-_{b''b'}(1-f_{k\alpha})\phi'_{ab''}\right]
}{i z-E_a+E_{b'}+E_{k\alpha}+E_{k'\alpha'}}.
\end{split}
\end{equation}
\end{widetext}
\end{subequations}

Eqs.~\eqref{generalPWw} and \eqref{generalPWphi} constitute the generalization of Eqs.~(10) and (11) in Ref.~\onlinecite{Pedersen05} to case with the counting field; they are the main formal result of our paper and the starting point for the following studies. Analogously to the mean current,\cite{Pedersen05,*PedersenPhD08,Konig:PRL96,*KoenigPhD98} charge conservation can be proven for all stationary cumulants using the method described in Ref.~\onlinecite{Emary:JPCM11}.  Eq.~\eqref{generalPWphi} can be (at least formally) solved for $\phi_{ab}$ in terms of $w_{ab}(z,\chi)$ and substituted into Eq.~\eqref{generalPWw} to give a closed non-Markovian generalized master equation for the reduced density matrix $w_{ab}(z,\chi)$ only. Its evolution kernel $\hat{\mathcal{W}}(z,\chi)$ (see below) can then be used in the machinery of Refs.~\onlinecite{Flindt:PRL08,Flindt:PRB10} to produce the current noise and higher-order cumulants of the full counting statistics. For ease of notation, we will only explicitly demonstrate this general procedure on the following example.  

{\em Single-resonant-level model.} ---
We consider an archetypical model of spin-less electrons and a single level forming the system
$H_{\mathrm{system}}\equiv E_d|1\rangle\langle 1|$. The only non-vanishing $T_{ba}(k\alpha)$ is $T_{10}(k\alpha)=t_{k\alpha}$ and consequently $T_{10}^\pm=e^{\mp i\chi_\alpha/2} t_{k\alpha},\ \bar T_{10}^\pm=e^{\pm i\chi_\alpha/2}t^{*}_{k\alpha}$. Defining analogously to Ref.~\onlinecite{Pedersen05} (apart from the factor of $2\pi$) the quantity $B_\alpha(E) \equiv 2\pi\sum_{k} t^{*}_{k\alpha}\phi_{10}(z,\chi;k\alpha)\delta(E-E_{k\alpha})$ (with $\bar B_\alpha(E)\equiv B_\alpha^*(E)\Big|_{\chi\to-\chi\atop z\to\bar z}$) and introducing the conventional tunneling rates $\Gamma_{\alpha}(E)\equiv 2\pi\sum_{k}|t_{k\alpha}|^{2}\delta(E-E_{k\alpha})$ and $\Gamma(E)\equiv\Gamma_{L}(E)+\Gamma_{R}(E)$, we get from Eqs.~\eqref{generalPWw} and \eqref{generalPWphi} 
\begin{widetext}
\begin{equation}\label{SRL}
\begin{split}
& i z w_{00}(z,\chi) - i w_{00}(t=0,\chi)  =	\sum_\alpha e^{i\chi_\alpha/2}\int\frac{dE}{2\pi}
	\left[ B_\alpha(E) - \bar B_\alpha(E)\right],\\
& i z  w_{11}(z,\chi) - i w_{11}(t=0,\chi)  = 	-\sum_\alpha e^{-i\chi_\alpha/2}\int\frac{dE}{2\pi}
	\left[B_\alpha(E)-\bar B_\alpha(E)\right], \\
&\left(E-E_{d} + i z - \int\frac{dE'}{2\pi}\frac{\Gamma(E')}{E-E' + i z}\right) B_\alpha(E) = e^{-i\chi_\alpha/2} f_\alpha(E) \Gamma_\alpha(E)\left(
	    w_{00}(z,\chi) + \sum_{\alpha'}e^{i\chi_{\alpha'}/2}\int\frac{dE'}{2\pi} \frac{\bar B_{\alpha'}(E')} {E'-E - i z}\right) \\
           &\qquad\qquad\qquad\qquad\qquad\qquad\qquad\qquad\qquad -e^{i\chi_\alpha/2}[1-f_\alpha(E)] \Gamma_\alpha(E)\left(w_{11}(z,\chi) - \sum_{\alpha'}e^{-i\chi_{\alpha'}/2}\int\frac{dE'}{2\pi} \frac{\bar B_{\alpha'}(E')}
           {E'-E - i z} \right).
\end{split}
\end{equation}
\end{widetext}
We have verified\cite{ZedlerPhD11} that, for $\chi=0$, these equations coincide with the RTA obtained by the real-time diagramatics.\cite{KoenigPhD98} 

\begin{figure}
\includegraphics[width=\linewidth]{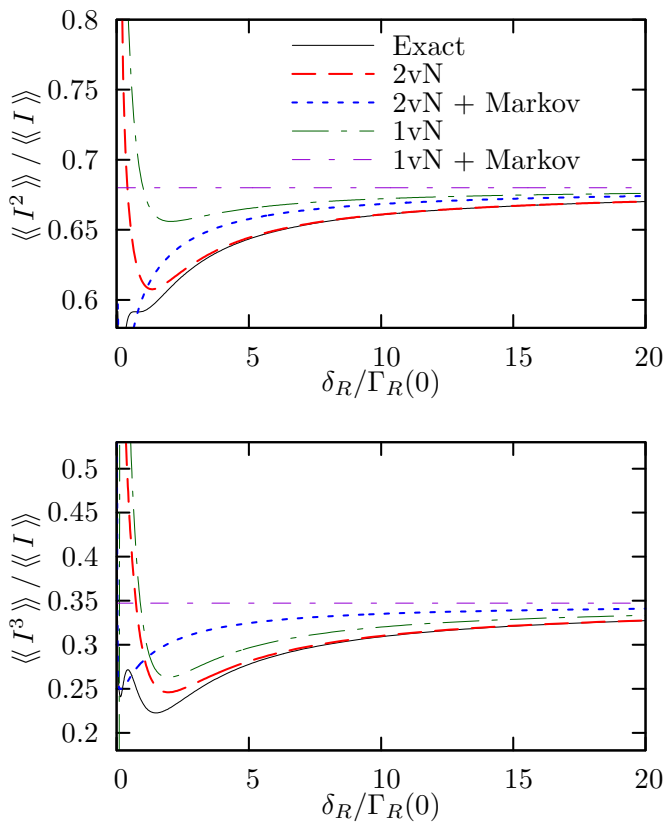}
\caption{(Color online) Second (top) and third (bottom) normalized cumulants of the current as functions of the width $\delta_R$ of the Lorentzian right tunneling rate measuring the degree of non-Markovianity. Compared are the exact solution (solid black line) and four tested approximations: 2vN/RTA (long-dashed red line), its Markovian limit (short-dashed blue line), the first-order von Neumann/standard GME (long-dot-dashed green line), and its Markovian limit serving as the reference (short-dot-dashed magenta line). \textsc{Parameters:} $\epsilon=\Gamma_R(0)=4\Gamma_L$.}\label{Fig:Lorentzian}
\end{figure}

From now on, we focus on a specific example of an infinite bias, i.e., $f_{L}(E)\equiv 1,\ f_{R}(E)\equiv 0$, with constant left tunnel rate $\Gamma_L(E)\equiv \Gamma_L$, and with the right tunnel rate having the Lorentzian energy dependence $\Gamma_R(E)\equiv 2|\Omega|^2\tfrac{\delta_R}{(E-E_R)^2+\delta_R^2}$.  This model has the advantage of being analytically solvable for both the exact solution as well as the RTA equations, yet it exhibits nontrivial non-Markovian dynamics. This model is equivalent\cite{Elattari00,Zedler09} to a noninteracting double dot with two electronic levels at $E_{d}$ and $E_R$
, mutually coherently coupled by the transfer amplitude $\Omega$, with the left level at $E_{d}$ coupled by $\Gamma_{L}$ to the left (filled) lead, and with the right level at $E_R$ coupled to the right (empty) lead by an energy-independent rate $2\delta_{R}$. The double dot model can be described exactly by a Markovian generalized master equation, which can be then projected onto the left dot with the level $E_{d}$ only resulting in an exact non-Markovian kernel $\hat{\mathcal{W}}_{\mathrm{exact}}(z,\chi)$. Its explicit form is too lengthy and cumbersome to be presented here so we just use it for the exact evaluation of the noise and skewness shown in Fig.~\ref{Fig:Lorentzian}.        

For the Lorentzian model, we can analytically solve the 2vN equations \eqref{SRL}, which is in itself remarkable and does not work at finite bias. This solution is obtained from the ansatz  $B_{R}(E)=b_{R}(E)/(E-\epsilon-i\delta_{R}),\ B_{L}(E)=b_{L}(E)$, where $b_{L,R}(E)$ has no singularities in the upper complex-$E$ half-plane. When this ansatz is inserted into the last of Eq.~\eqref{SRL} adapted to our example, one obtains, consistently with the assumed analytic structure in $E$, the following solution:
\begin{equation}\label{LorentzBs}
\begin{split}
B_L(E)& = 	e^{-i\chi_L/2}\Gamma_L \frac{w_{00}(z,\chi)+ie^{i\chi_R/2}\frac{\bar b_R(\epsilon-i\delta_R)}{E+i z -\epsilon+i\delta_R}}
{E+i z-\Sigma(E+i z)},\\
B_R(E)& =	-\Gamma_R(E)	\frac{	w_{11}(z,\chi)e^{i\chi_R/2}-i\frac{\bar b_R(\epsilon-i\delta_R)}{E-\epsilon+i\delta_R+i z}}{E+i z-\Sigma(E+i z)},
\end{split}
\end{equation}
with $\epsilon\equiv E_R-E_d$ and $\Sigma(E)= -i \Gamma_{L}/2+|\Omega|^{2}/(E-\epsilon+i\delta)$, the self-energy due to coupling to the leads corresponding to the sum of the tunneling rates. Using 
the relation $b_R(\epsilon+i\delta_R) =\lim_{E\to\epsilon+i\delta_R}[E-(\epsilon+i\delta_R)]B_R(E)$ stemming from the above ansatz for the determination of $b_R(\epsilon+i\delta_R)$ from the second of Eq.~\eqref{LorentzBs} and inserting the resultant form of Eq.~\eqref{LorentzBs} into the first two lines of Eq.~\eqref{SRL}, we finally arrive at the non-Markovian kernel in the 2vN approximation, 
\begin{equation}\label{RTA}
\hat{\mathcal{W}}_{\mathrm{2vN}}(z,\chi)=\begin{pmatrix}
	  -\Gamma_L & e^{i\chi_R}\frac{2|\Omega|^2\gamma(z)}{\epsilon^2+\gamma(z)[\gamma(z)+|\Omega|^2/(\delta_R+ z/2)]} \\
	  \Gamma_L e^{-i\chi_L} & -\frac{2|\Omega|^2\gamma(z)}{\epsilon^2+\gamma(z)[\gamma(z)+ z+|\Omega|^2/(\delta_R+ z/2)]}
	\end{pmatrix},
\end{equation}
with $\gamma(z)\equiv \delta_R+\Gamma_L/2+ z$.

For zero counting fields, this kernel is identical to the exact one, $\hat{\mathcal{W}}_{\mathrm{2vN}}(z,0)=\hat{\mathcal{W}}_{\mathrm{exact}}(z,0)$, with clear consequence that the time-dependent occupations and stationary current are also exact in accordance with previous findings.\cite{Pedersen05,Pedersen07,PedersenPhD08,Zedler09} The non-equilibrium rates are captured correctly including the effects of broadening of the level $E_{d}$ due to the coherent coupling to the leads. However, already at first sight the counting field is accounted for in a rather primitive manner analogous to the lowest-order sequential tunneling model, i.e., the exact rates are just multiplied by the exponentials with counting fields (compare with Eq.~(19) of Ref.~\onlinecite{Zedler09}). For non-Markovian kernels, this prescription is potentially problematic\cite{Flindt:PRB10} and the exact kernel indeed contains the counting fields also in the denominators of the expressions for the rates ensuring consistency. Therefore, the 2vN kernel with counting fields is not exact and the noise and higher-order cumulants it yields are not correct. 

This is explicitly demonstrated in Fig.~\ref{Fig:Lorentzian}, where we compare the exact solution for the noise and third cumulant with the 2vN/RTA and its Markovian limit obtained by using $z\to 0$ in the kernel \eqref{RTA}.\cite{Flindt:PRB10}  We also compare with the first-order von Neumann approach (1vN) equivalent to the standard lowest-order GME described in Sec.~IIIA of Ref.~\onlinecite{Zedler09}, and its Markovian limit. We fix the right tunnel rate $\Gamma_{R}(0)$ by adjusting $\Omega$ while changing the parameter $\delta_{R}$ controlling non-Markovian behavior and observe the performance of various approximations for increasing degree of memory with decreasing $\delta_{R}$. The Markovian limit of the 1vN is left constant by this procedure and serves as a reference, while all other solutions respond to the change of $\delta_{R}$. Obviously, none of the approximations reproduces the exact solution for strong-enough non-Markovian dynamics; all approximations perform quite badly for strong memory at $\delta_{R}\sim\Gamma_{R}(0)$ for both cumulants \new{with extreme errors in the non-Markovian versions. These errors are typical for non-Markovian master equations which have no Lindblad form and hence do not guarantee the conservation of probability \cite{Zedler09}.} However, in the intermediate memory regime $\delta_{R}\gtrsim\Gamma_{R}(0)$ \new{the non-Markovian version of} 2vN/RTA is clearly by far the best approach.               

{\em Conclusion.} We have extended the second-order von Neumann approach for GME kernels to the case with counting field which has enabled us to study its performance in the evaluation of the electronic current noise and higher cumulants. We have shown on an analytically solvable example of a single-level junction with structured coupling to the leads that 2vN is {\em not} exact for the noise and higher-order cumulants and that it may fail significantly for highly non-Markovian dynamics. However, for intermediate degrees of memory, 2vN seems to perform the best out of considered standard approximations.  Thus, when supplemented with efficient numerical implementation, it would be a method of choice for evaluation of noise and full counting statistics for systems with strong interplay between correlations and coherence.      

{\em Acknowledgments.} We thank  O.~Karlstr\"om,  P.~Samuelsson, and A.~Wacker for
useful discussions. This work was supported by DFG Grants BR 1528/7-1, 1528/8-1, SFB 910, GRK 1558, the Heraeus foundation, and the DAAD, and  by the Czech Science Foundation via Grant  No.~204/11/J042 and the Ministry of Education of the Czech
Republic via the research plan MSM 0021620834 (T.~N.). 

\bibliography{RTA}

\end{document}